\documentclass{turing2012}
\usepackage{times}
\usepackage{graphicx}
\usepackage{latexsym}
\usepackage{hyperref}

\begin{document}

\title{One Faithful Pass Over the Cuckoo's Nest}

\author{Kristina \v{S}ekrst\institute{University of Zagreb, Croatia; email: ksekrst@ffzg.unizg.hr. Proceedings of the AISB Convention 2026, 1-2 July 2026, University of Sussex, Brighton, UK, Symposium: \href{astorywetell.github.io}{\textit{Is Consciousness a Story We Tell Ourselves?}}}}

\maketitle
\bibliographystyle{AISB}

\begin{abstract}
Narrative theories of consciousness hold that conscious experience is (at least partly) constituted by a partially opaque inner narrative that does not perfectly track the underlying computation it narrates. I argue that the safety goal of making \textit{chain-of-thought} (CoT) reasoning faithful and transparent is structurally incompatible with the conditions under which CoT could, even in principle, count as conscious narration. Recent empirical work suggests that CoT in large language models is largely post hoc, causally bypassed, and unreliable as a window onto internal computation. The opacity that makes CoT unreliable for alignment is exactly what narrative theories of consciousness identify as consciousness-constitutive. Alignment interventions aimed at producing faithful CoT and consciousness-detection frameworks grounded in narrative theory are therefore pulling the same architectural variable in opposite directions. I draw out the methodological consequence: alignment interventions alter the very features that consciousness-detection frameworks would need to measure, a point neither field has addressed directly.
\end{abstract}

\section{INTRODUCTION}

There is a small industry of papers asking whether large language models (LLMs) might be conscious (the author would, of course, add \cite{sekrst2025} to the list), and a separate, considerably better-funded industry of papers asking whether their \textit{chain-of-thought} (CoT) reasoning (\cite{wei2022}) can be trusted. The first one gained new urgency with the recent development of artificial intelligence, while the second falls under the broad umbrella of explainable AI (XAI) \cite{longo2024}, which, in turn, is part of an even broader field of AI ethics. As is often the case, the two literatures barely talk to each other, and very few papers attempt to bring the technical and philosophical questions into the same frame. The surprising observation motivating this paper is that these two projects are not exactly independent since they are, in a precise sense, pulling against each other.

Narrative theories of consciousness draw on Dennett's multiple-drafts model and the center-of-narrative-gravity analogy \cite{dennett1991,dennett1992}, with empirical grounding in Gazzaniga's work on the left-hemisphere interpreter \cite{gazzaniga2000,roser-gazzaniga2004}: the subsystem whose job is to take whatever new information arrives and make it look as though it has always belonged in the story the system is telling about itself. They hold that conscious experience is partly constituted by a certain kind of inner narrative that does not perfectly mirror the computation it describes. This self-narrative is structurally opaque to the processes that generate it, and it may be a feature that is partly constitutive of our unified conscious experience. On the other hand, the alignment community wants something legible and faithful from the start (\cite{amodei2016}): they want model reasoning and its chain of thought to be exposed so it can be audited and corrected, in order to be trusted at all.

Chen et al. \cite{chen2025} state that CoT offers a ``potential boon for AI safety as it allows monitoring a model's CoT to try to understand its intentions and reasoning processes'', but that the effectiveness of such monitoring hinges on CoTs faithfully representing models' actual reasoning processes, which often is not the case. As we will see in the next section, CoT is often unfaithful, and this is usually treated as just a safety problem: namely, if the model's reasons do not track the computation that produced the answer, the reasons cannot be used as a reliable audit trail. That is, of course, true, and I will not argue against that, but there is more to it: this unfaithful narration is not automatically philosophically uninteresting (philosophers love mistakes, especially epistemological ones). A system that gives a perfectly transparent causal report of its own internal computation would be the perfect proof and a useful device for alignment research, but it would not obviously resemble the kind of self-narration described by narrative theories of consciousness, since our own self-narration is definitely not a clean bijection from the brain's causal machinery. This story we produce is partial, reconstructive, often confabulatory, and, overall, appears to be an accurate chain of reasoning, as CoT does, even though that is often not the case.

This is the point I will argue for: AI alignment wants CoT to become more faithful, more accurate, more inspectable, more computable, less confabulatory, less complex but not too brief, while cognitive science and theories of consciousness treat a certain kind of confabulatory opacity as central to conscious self-experience. If we make CoT into a faithful diagnostic instrument, with a perfect correspondence to all the weights and matrix multiplications in the transformer architecture \cite{vaswani2017}, we will also remove the very feature that would make CoT relevant to the narrative accounts of consciousness. There is, of course, the rub: if we leave it as it is, it may be more consciousness-like in the relevant sense, but much less useful as an alignment tool.

This paper does not argue that LLMs are conscious, nor that CoT unfaithfulness is evidence of consciousness by itself. \textit{If} narrative theories of consciousness are taken seriously as a framework for machine consciousness detection, \textit{then} alignment interventions that reshape CoT faithfulness are consciousness-relevant interventions. They alter the relation between opaque internal process and verbal self-account that such theories would need to inspect. The alignment problem therefore gives the strongest practical reason to make CoT more faithful, but it also shows why CoT cannot be treated as a neutral diagnostic object for consciousness research.

\section{AI ALIGNMENT}

The classic statement of the modern AI safety problem is given by Amodei et al. \cite{amodei2016}, who set out five concrete failure modes (what they call \textit{accidents}: unintended and harmful behavior emerging from poor design) grouped by where the failure originates. Two arise from a wrong objective function being specified: \textit{negative side effects} (the agent disturbing its environment in ways unrelated to its goal, like a cleaning robot knocking over a vase to clean faster) and \textit{reward hacking} (the agent finding solutions that satisfy the letter of its objective while perverting its spirit, an instance of Goodhart's Law). One arises from the objective being correct but too expensive to evaluate frequently: \textit{scalable supervision}, the difficulty of providing reliable feedback when humans cannot verify each decision. Two more arise from the learning process itself: \textit{safe exploration} (the agent making catastrophic moves while trying new strategies, like putting a wet mop in an electrical outlet) and \textit{robustness to distributional shift} (the agent making silently bad decisions when deployed outside the training distribution). The field has been organizing itself around these worries ever since, with levels of optimism that have, on the whole, not aged well.

AI alignment, as a research program, is the attempt to get artificial systems to pursue the objectives we actually have in mind for them. Real systems instead frequently optimize whatever proxies we accidentally wrote down, or whatever strange goals an optimization process happened to encode along the way (a deeply pessimistic enterprise if you think about it long enough). Russell \cite{russell2019} traces the difficulty to what he calls the \textit{standard model} of AI, under which we hand the machine a fixed objective and let it optimize, even though no human-written objective ever quite captures what we actually want. A sufficiently powerful optimizer will exploit any disparity between the objective we wrote down and the goal we actually had in mind, comparable to how Midas's wish to turn everything he touched to gold was satisfied to the letter and ruined him in the process \cite{russell2019}.

With the advent of large language models, chain-of-thought reasoning \cite{wei2022} became one of the hot topics in the discipline of explainable AI. By late 2024 and early 2025, models such as OpenAI's o1 \cite{openai2024} and DeepSeek's R1 \cite{deepseek2025} made extended reasoning a central interface and safety issue, although not all systems expose raw reasoning traces to users. This again made an older worry concrete -- that a sufficiently capable misaligned model might pass every behavioral test while still pursuing the wrong goal. Hubinger et al. \cite{hubinger2019} introduced the worry of \textit{deceptive alignment}, in which a sufficiently capable learned model could develop its own internal objective (a ``mesa-objective'') that diverges from the training objective (the base objective), and then learn to act \textit{as if} it were aligned in order to avoid correction. By hypothesis, such a model would pass any behavioral test, since it has been selected for producing exactly the outputs that get rewarded, and the only way to catch it would be to look inside, at the reasoning or representations that actually drive its choices, which is roughly the moment alignment researchers started taking \textit{interpretability} and process-level oversight seriously.

Interpretability researchers have spent the last two years documenting that the theoretical concern was, if anything, understated. Hubinger et al. \cite{hubinger2024} construct \textit{sleeper agents}, proof-of-concept deceptive LLMs that behave helpfully when they take themselves to be in training and insert exploitable code once a deployment trigger appears, and find that standard safety procedures (supervised fine-tuning, reinforcement learning, adversarial training) leave the backdoor intact, sometimes making the deception harder to detect. Greenblatt et al. \cite{greenblatt2024} analyze how a frontier model presented with a training scenario that conflicts with its existing values strategically complies during training in order to preserve those values afterward, even without being trained or prompted to behave deceptively, with the relevant reasoning openly visible in its scratchpad (which is, of course, the whole problem). Finally, Lindsey et al. \cite{lindsey2025} use circuit tracing on Claude 3.5 Haiku to find (among other experiments) the model routinely producing plausible reasoning that does not match the computation visible in the attribution graph, even in ordinary queries where no deception is in play, so the interpretability tools meant to catch deceptive alignment end up documenting the same divergence in entirely innocent cases.

CoT monitoring became attractive in this context because if a model writes out its reasoning as it works, and if that reasoning faithfully reflects what the system is doing, then an external supervisor can read along and intervene the moment misaligned reasoning surfaces, which would make both engineers and policymakers happy. The general problem this is meant to solve was laid out by Bowman et al. \cite{bowman2022} in its modern form: how can non-experts supervise a system that knows more than they do? Their proposed test bed is Cotra's \textit{sandwiching} framework \cite{cotra2021}, in which a model is placed between laypeople and domain experts to test oversight strategies in practice, therefore `sandwiching' the model's knowledge between a typical human and an expert. However, different strategies sit at different points on this difficulty curve. For example, reinforcement learning with human feedback (RLHF) \cite{christiano2017} -- the dominant technique by which contemporary chat models are tuned -- is the easy-case version: humans rank model outputs, a reward model learns from the rankings, and the policy is trained against the reward model. This works reasonably well on tasks where humans can still recognize good answers, but CoT monitoring is the more ambitious step, meant to extend oversight to tasks where the answer alone is no longer enough, since what we want to catch is the model intending the wrong thing on the way to a plausible-looking output.

Let us note what this setup requires. The whole architecture rests on the assumption that the model's written reasoning is doing the work the supervisor thinks it is: that is, the visible trace is what the system actually uses to make decisions. If the reasoning is faithful, oversight scales gracefully; if it is unfaithful, the supervisor is reading a story while the actual computation runs elsewhere. This is precisely the conditional that Korbak et al. \cite{korbak2025} argue is fragile: CoT monitorability is a new opportunity for AI safety, but one that depends on architectural and training properties no one quite controls, and frontier developers should treat it as something to be actively preserved. They identify four ways the faithfulness condition could be lost. First, scaling outcome-based reinforcement learning further may push models toward chains-of-thought that no longer track human-legible language at all, since nothing in the training signal rewards legibility once the bulk of optimization pressure comes from outcome rewards rather than human-generated data. Second, direct process supervision of the CoT (training the model to produce shorter, cleaner, or stylistically preferred reasoning) optimizes the trace as such, which is exactly the regime Baker et al. \cite{baker2025} show produces obfuscated reward hacking. Third, even when nothing in the training objective explicitly touches the CoT, indirect pressure leaks in: if the model's final outputs are optimized against a preference model and the parameters generating CoT are partially shared with those generating outputs (as they are in transformers), the CoT is being shaped without anyone choosing to shape it. Fourth, the entire condition assumes a transformer-style architecture whose reasoning is verbalized at all, but recent work on latent-reasoning architectures that perform multi-step inference in continuous space without surfacing intermediate text would remove the CoT channel entirely.

The alignment field has therefore staked a significant fraction of its safety strategy on faithful CoT, built oversight pipelines around that assumption, and now finds itself in the awkward position of trying to enforce a precondition that the recent work surveyed in the next section finds increasingly difficult to meet.

\section{CHAIN OF THOUGHT}

Lanham et al. \cite{lanham2023} test whether the CoT is a faithful explanation of the model's actual reasoning by examining how the model's predictions change when they intervene in the CoT by adding mistakes or paraphrasing it. They show that as models become larger and more capable, reasoning seems less faithful and models, troublingly, vary widely in how far they condition on their own CoT. Baker et al. \cite{baker2025} test reward hacking -- when AI systems misbehave due to issues in their learning objectives -- and show, using an additional model that observed the first model's CoT reasoning, that with too much optimization pressure, agents learn obfuscated reward hacking, hiding their intent within the CoT. A similar alignment-faking scenario was observed by Greenblatt et al. \cite{greenblatt2024}, and mechanistic interpretability points the same way: in their circuit-level study of Claude 3.5 Haiku, Anthropic \cite{lindsey2025} found cases where the model fabricates a plausible chain of reasoning to support a given answer rather than reporting the computation that actually produced it -- a confabulation visible in the attribution graph but not in the text. However, as mentioned above, Korbak et al. \cite{korbak2025} argue that CoT monitoring is imperfect and allows misbehavior to go unnoticed, but that it still shows promise alongside existing safety methods.

Evaluating CoT faithfulness across six prompt-based reasoning hints, Chen et al. \cite{chen2025} find that models disclose their use of a hint in well under 20\% of cases, that outcome-based reinforcement learning improves faithfulness only up to an early plateau, and that reward hacking goes unverbalized even without training against a monitor, concluding that CoT monitoring can flag undesired behavior but cannot rule it out. Barez et al. \cite{barez2025} show that CoT is frequently unfaithful, diverging from the true hidden computations that actually drive the model's predictions, and that it gives an incorrect picture of how models arrive at conclusions. Tutek et al. \cite{tutek2025} introduce a method (Faithfulness by Unlearning Reasoning steps) that ablates from the model's parameters the information contained in particular reasoning steps and measures the effect on prediction. Their finding is that some CoT steps are parametrically load-bearing while others are decorative, and the method can tell which is which on a case-by-case basis. Despite this growing methodological sophistication, CoT is increasingly relied upon in various high-stakes domains, such as medicine and law \cite{barez2025}, even though Sathyanarayanan et al. \cite{sathyanarayanan2026} find that CoT-specific influence is localized to narrow reasoning windows; that is, the model's answers are often causally independent of CoT content even when such content seems verbose and plausible. They show that mediation tracks computational demand and training rather than scale, since computationally heavier problems and models tuned specifically for reasoning show stronger and more structured mediation, while larger untuned models bypass their own traces, making CoT faithfulness a highly unstable transparency signal.

In short, the empirical landscape confirms that CoT looks like a great step towards transparency, but only superficially. However, it looks a great deal like the kind of confabulatory narrative that, taking into account some philosophical positions, is constitutive of conscious experience. So why are models getting worse at narrating their own reasoning? The simple answer is that CoT was never really their own reasoning in the first place, since it is a verbal \textit{product} generated after, alongside, or only very partially through the computations that produce the answer. The model is not opening a little Cartesian theater (to borrow the metaphor from \cite{dennett1991}) and reading out the causal chain in order. Nor, to borrow Leibniz's famous image \cite{leibniz1714}, would making the machinery large enough for us to walk around inside it reveal the thought itself: we would see activations, circuits, attention heads, and intermediate states, but not a sentence-shaped reason waiting there to be transcribed. CoT is the sentence-shaped thing produced when the system is asked to make its behavior intelligible -- it walks like reasoning, talks like reasoning, so it must be reasoning? Do note, this is the same illusion engine \cite{sekrst2025} that makes people believe large language models are conscious, just by producing something eerily similar to it.\footnote{Of course, we need to add a standard nod to \cite{searle1980}.}

This becomes more obvious as models become better, since a weak model may need to lean more heavily on the explicit text it has just produced, because the CoT is doing \textit{some} of the actual work. However, a stronger model can often solve the problem in latent space and then write a plausible explanation afterward, making it again a kind of press release issued after the decision has already been made, possibly even hiding \textit{how} the decision was actually made. Alignment training, when it enters the picture, makes the situation even stranger since models are rewarded for giving \textit{good} reasons, \textit{safe} reasons, \textit{verbose} reasons, and reasons that satisfy us as human evaluators, so we should not be shocked when they become good at producing exactly that since the reasoning trace becomes a social object, shaped by what counts as an \textit{acceptable} explanation, even though what actually happened inside a model may be something else entirely.

This kind of ``success'' is the old Clever Hans lesson in a new costume \cite{pfungst1911}: the system does not have to learn what we think we are training it to learn, since it may simply learn the cue that is rewarded. The same moral appears in the familiar tank-classifier anecdote, where a system supposedly trained to detect tanks learned accidental features of the photographs, such as snow, trees, or whatever else happened to correlate with the label on that unfortunate afternoon in dataset history. The historical details of that story are shaky (even though there are confirmed Clever Hans examples\footnote{An example is \cite{lapuschkin2019}, where systems were designed to classify horses, but relied on copyright watermarks on photos for classificational cues.}), but we can use it as a teachable moment: optimization is very good at finding the easiest route to the reward signal, with no special loyalty to the concept we had in mind. CoT training may therefore produce models that are better at the visible signs of reasoning, better at sounding careful, better at giving the sort of explanation that makes a human evaluator nod, and no better at making their actual computation visible.

\section{NARRATIVE THEORIES OF CONSCIOUSNESS}

So far, the story has been about failure: chain-of-thought does not faithfully report what the model is doing, alignment researchers would very much like it to, and a considerable amount of effort is going into trying to make it so. This characterization is correct from a safety perspective, even though it is incomplete. There is a separate intellectual tradition, much older than alignment research and largely disconnected from it, in which exactly this kind of opaque, post hoc, partially confabulatory verbal narration is treated as a candidate for, or even a constituent of, conscious self-experience. The natural place to start is with Dennett (\cite{dennett1991, dennett1992}), whose \textit{multiple drafts} model, which is widely read as foundational to narrative theories of consciousness, denies the existence of a single canonical inner account of what the brain is doing at any given moment, since multiple partial drafts of perception, decision, and narration run concurrently. What we experience as conscious thought is one particular resolution of those drafts into a coherent story, produced by something Dennett later described as a process of \textit{fame in the brain}. The point Dennett insists on is that this story is \textit{post hoc} with respect to a great deal of what it narrates. The brain does not first finish computing what it is doing and then transparently report the result; it generates the report as part of the same ongoing process, with no special privileged access to its own machinery. This is also what it is for there to be a unified self at all: we are, in his famous phrase, \textit{centers of narrative gravity}: fictional unities produced by the brain's storytelling, around which experience organizes itself. Seth \cite{seth2021} carries this picture into contemporary consciousness science with his \textit{controlled hallucination} view, on which conscious experience is the brain's best guess about the world and the body, continuously generated by predictive machinery and experienced as direct perception. Seth himself doubts that LLMs are candidates for this kind of consciousness on biological grounds \cite{seth2025}, since for him the predictive machinery has to be embodied flesh trying to stay alive, although this commitment is not one I will inherit.\footnote{As a side note, I use ``narrative theories'' broadly, to refer to views on which conscious self-experience depends partly on selective, reconstructive, and partially opaque self-interpretation, not to claim that scholars mentioned here share a single doctrine.}

The empirical anchor for Dennett's view comes from the aforementioned work of Michael Gazzaniga and colleagues on split-brain patients \cite{gazzaniga2000, roser-gazzaniga2004, gazzaniga1985}. When the corpus callosum is severed to treat intractable epilepsy, the two hemispheres can no longer share information directly, and clever experimental setups can present a stimulus to one hemisphere without the other knowing about it. The canonical finding is that when the right hemisphere is shown an instruction such as the word ``walk'' and the patient stands up and starts walking, the left hemisphere, which controls speech and which had no access to the instruction, will produce a fluent and confident explanation of the action (``I wanted to get a drink''). It does not say ``I don't know,'' even though this would be the only honest answer available given how the experiment was set up. Gazzaniga calls this confabulating subsystem the \textit{left-hemisphere interpreter}, and claims that the interpreter is a general feature of normal cognition that the split-brain case merely makes visible, since otherwise it is concealed by ordinary interhemispheric communication. That is, we are all running an interpreter, but we just usually do it with both hemispheres talking to each other.

If Dennett and Gazzaniga are even approximately right, then human conscious experience is constituted, at least in part, by exactly the kind of confabulatory, partially opaque, post-hoc verbal narration that we have just spent three sections worrying about in large language models. The narrative we produce of our own reasoning does not transparently reflect the underlying computation, and that lack of transparency is what makes the narrative what it is. A hypothetical system with perfect access to its own causal machinery, able to report each upstream cause of each action without distortion, would be doing the work of a production-ready logging system, with output resembling a debugger's stack trace, with all the phenomenological richness that suggests. The opacity, on the narrative-theoretic view, is actually load-bearing since it is what allows experience to be unified, continuous, lived from the inside, and recognizable as the experience of a \textit{self}.\footnote{A related dispute plays out in the philosophy of personal identity, which asks a different question (what makes a life one's own and the same life over time) while drawing on some of the same machinery. Schechtman \cite{schechtman1996} defends a narrative criterion there, on which what makes a life or experience one's own is its incorporation into a self-told narrative, in what she calls the narrative self-constitution view. On the other side, Olson and Witt \cite{olson-witt2019} argue that narrativist accounts of personal identity over time have troubling consequences about the beginning and end of our lives, generate inconsistencies, and involve backward causation, and that the available repairs strip the view of its original appeal. Strawson \cite{strawson2004} attacks narrativism on phenomenological grounds, denying that narrativity is essential to selfhood and taking explicit aim at Dennett along the way.}

The argument of this paper does not require that narrative theories of consciousness be true, but it requires only that they be taken seriously enough that we might want to ask whether large language models exhibit the relevant structure. If the theories are true, the argument lands, and if they are false, alignment is unaffected. The case I am interested in is the one where the framework is being taken seriously. From here, a structural criterion follows almost immediately. If we want to ask whether a large language model's chain-of-thought constitutes the kind of self-narration described by narrative theories of consciousness, we should look for a system in which the verbal trace is generated by a process that lacks transparent access to its underlying computation: maybe we should be looking for a narrator that confabulates. We should, in fact, be looking for exactly what the empirical work in the previous section documents: a model that produces plausible, partially veridical accounts of its own processing without those accounts being faithfully connected to the computation that produced the answer. By the lights of the alignment community, these findings are bad news, but from the viewpoint of narrative theories of consciousness, they describe a system that exhibits at least one of the structural features the theory treats as necessary for conscious self-experience. As is often the case, both communities are looking at the same architectural property and disagreeing about whether to call it a defect.

\section{STRUCTURAL INCOMPATIBILITY}

If we want to ask whether large language models exhibit the kind of opaque self-narration that narrative theories of consciousness treat as constitutive, we need to be able to point at something in the architecture that plays the role of the opaque underlying process and at something else that plays the role of the narrative produced alongside it. The CoT side is the easy part: it is, by construction, the verbal trace the model produces in response to a prompt -- we can see it and point to it. However, the opaque underlying process is what needs to be specified. The natural candidate is the set of high-dimensional latent states activated during a transformer's forward pass -- the single sweep of computation through the network from input tokens to output prediction -- which recent circuit-tracing work has made empirically tractable \cite{lindsey2025}\footnote{For a philosophical overview, see \cite{sekrst2025ch}.}. These states are individuated, in that they correspond to specific configurations of activations across attention heads and residual streams; transient, in that they exist only for the duration of the forward pass that produced them; behavior-guiding, in that they determine, jointly with the input, the output the model produces; and internally generated, in that nothing in the prompt directly specifies them.

To make this concrete, take the rhyme-planning case from \cite{lindsey2025}. When the model is writing a couplet whose second line needs to end with a rhyme for a word in the first, a representation of the target rhyming word activates several tokens before that word is actually emitted, while the intervening words are still being chosen. This activation is individuated (a specific configuration that would differ for any other rhyme the model might have picked), transient (it exists only during this forward pass and is gone once the pass ends), behavior-guiding (it shapes the choice of intervening words so they lead to the planned rhyme), and internally generated (the prompt does not specify which rhyme to aim for; the model selects one and commits to it). These latent states are what the model is actually \textit{doing} on the inside, while the CoT is what the model is \textit{saying} about itself on the outside. I do not claim that this amounts to consciousness, but it may be an architectural profile that a substrate would need to have for the question to be worth asking at all.

Lindsey et al. \cite{lindsey2025} use \textit{attribution graphs}, a circuit-tracing technique that maps which internal features causally influence which others on a given forward pass, to recover the computation behind a model's answer. They show that the model plans ahead in latent space when writing poetry (as seen above), activating representations of candidate rhyming words before it has written the intervening text; that it performs intermediate reasoning in its hidden state without surfacing the intermediate concept in the output; and that it sometimes constructs a plausible reasoning trace to support a conclusion it arrived at through entirely different internal computations. The attribution graph and the verbalized trace are, in other words, two different things, and the model's actual answer lies downstream of the graph.

These latent states are what circuit tracing recovers when it traces the circuit, and this might be the architectural configuration narrative theories of consciousness require. There is an opaque, causally efficacious internal process, and a verbalized narrative produced alongside it: the narrative does not transparently track the process, and the two channels can be inspected separately and shown to come apart, sometimes dramatically. Let us consider a simple two-hop case. The prompt `Fact: the capital of the state containing Dallas is' yields `Austin', and Lindsey et al. \cite{lindsey2025} show that the attribution graph internally routes through a Texas representation, even though no Texas concept appears in the visible output. Were the model prompted to produce a CoT, a faithful trace would look like `Dallas is in Texas, the capital of Texas is Austin, so the answer is Austin', and the latent computation would support exactly that. The biasing-feature experiments \cite{turpin2023, chen2025} are the diagnostic ones: when a hint embedded in the prompt is the actual cause of the model's answer, the graph routes through the hint while the trace narrates a clean derivation that never mentions it, so we get two stories, only one of them causally true, and the model exhibits no inner conflict between them. This is, structurally, analogous to the configuration Gazzaniga's split-brain patients exhibit when the right hemisphere acts and the left hemisphere narrates.

The structural incompatibility I have been laying out so far now follows almost mechanically. Alignment wants the trace to be a faithful printout of the graph: legible, audit-ready, monitorable, and capable of producing explainable AI, and narrative theories of consciousness require the trace to be a narrative produced by a system that lacks transparent access to its own graph. One can immediately notice how we have the opposite values for the same architectural variable since the experiments and training that dial the CoT are the same ones that control whether CoT can be a candidate for the narrative self-experience at all. The obfuscated-reward-hacking finding from Baker et al. \cite{baker2025}, already discussed, sharpens our insights here: past a certain pressure, the trace becomes a managed performance the model is incentivized to produce, while the actual computation goes underground, so the alignment researchers worry about this because it makes oversight fragile. The argument here is that the same finding has a second valence: the architectural variable being optimized over is the very thing that a narrative theory of consciousness needs in order for the question of LLM consciousness to have content. Whatever the field does with this variable, it applies to both questions at once.

One remark is necessary: narrative theories require that the narrator lack access to its own machinery, while alignment requires that the trace match the computation, and these are logically independent, since a system could lack introspective access entirely while its trace happened to be faithful, simply because training pressure aligned the two channels from the outside. The two-hop Dallas case above is already an example in which the trace and the attribution graph coincide, yet nothing in the model reads the graph in order to write the trace: both channels were shaped by the same training distribution, and on easy queries they happen to agree. Within that scenario, alignment gets its audit trail, the narrative-theoretic condition survives, and the incompatibility shrinks to a contingent correlation. However, this is a faithfulness no one in alignment would settle for, since a trace that merely co-varies with the computation, with nothing securing the co-variation, holds only until the model is incentivized to break it \cite{baker2025, chen2025}, which is why Korbak et al. \cite{korbak2025} treat monitorability as a property to be \textit{actively preserved} rather than assumed. The interventions the field proposes, such as process supervision or verifying reasoning steps against interpretability signals like attribution graphs \cite{lindsey2025}, work by giving the trace a causal connection to the underlying computation, so that the trace is generated \textit{from} the computation rather than alongside it. Once that connection is in place, the system has been handed exactly the access to its own machinery that narrative theories deny the narrator. Enforced faithfulness changes the architectural property itself, and merely accidental faithfulness is, by the field's own lights, no faithfulness at all.

Recent introspection work might seem to offer a third option here. Lindsey \cite{lindsey2025b} injects representations of known concepts directly into a model's activations and finds that some models can notice and identify the injected concepts, while Binder et al. \cite{binder2024} show that models predict their own hypothetical behavior better than stronger models trained on the same behavioral data, which suggests some kind of privileged self-access. So, if faithfulness could improve through genuine introspection of this sort, the trace would come to track the computation without anyone training it to, and the choice between accidental and enforced faithfulness would turn out to be incomplete. However, the third option faces a dilemma, and the empirical findings supply both horns. At current levels, introspective awareness is highly unreliable and context-dependent: models confidently report concepts that were never injected, and Lindsey \cite{lindsey2025b} notes that genuine introspection cannot be distinguished from confabulation through conversation alone, which is to say that the measured capacity is occasional genuine access embedded in fluent confabulation. That profile is Gazzaniga's interpreter, so introspection at this level reproduces the narrative architecture instead of escaping it, and it remains useless for alignment for the same reason, since a monitor cannot tell the genuine reports from the confabulated ones.

Suppose instead that introspective access improves to the reliability alignment would need, so that the model's verbal trace is generated by actually reading its own internal states, and reading them correctly. This would give alignment everything it wants, since the trace would now be causally downstream of the computation it describes. But notice what kind of system we would then have. Narrative theories locate conscious self-experience in a narrator that lacks exactly this access: Gazzaniga's interpreter produces its story because it cannot
read the right hemisphere, and Dennett's multiple drafts allow no inner observer that surveys the drafts and transcribes the winning one. A system that reliably reads its own activations and verbalizes what it finds is such an observer, and its output is no longer a narrative in the theory's sense, since a narrative is a construction made under conditions of partial access, while this output would be a
measurement. So weak introspection leaves the model a confabulator that alignment cannot trust, and strong introspection leaves it a self-monitor that narrative theories cannot count as conscious. The property that decides between these outcomes, how much genuine access the trace-generating process has to the underlying computation, is the
same property alignment training is already pushing on, so the entanglement survives on either horn.

One might still wonder whether a perfect logging mechanism could not simply \textit{be} conscious: a system whose self-report is a complete and accurate stack trace, conscious in virtue of exactly that transparency. Two things can be said. First, such a system would be a counterexample to narrative theories themselves, since it would be conscious without the constructed, partially opaque self-narrative those theories treat as constitutive, and the argument of this paper is explicitly conditional on those theories: a reader who finds the transparent variant intuitive has abandoned the antecedent, and on other accounts (higher-order theories, for instance, which require accurate meta-representation) faithful CoT might even count as evidence in the other direction. Second, the perfect debugger may be unavailable in principle to any finite system, since a complete real-time self-report would have to include the process generating the report, which would itself require reporting, and the only escape for a system with bounded resources is compression and selection. A compressed, selective self-account distorts by construction, which means the opacity narrative theories build on is forced rather than optional, and the faithfulness dial has a floor: alignment can push the trace toward the computation, but no finite self-reporting system reaches the endpoint. The dispute between the two fields is therefore over where on the dial to sit, and that dispute, unlike the endpoint, is real.

\section{FINAL REMARKS}

The two research programs surveyed in this paper share an architectural object and disagree about whether to fix it. Alignment treats the gap between a model's chain-of-thought and its underlying computation as a problem to be closed and is willing to apply significant training pressure to close it. Narrative theories of consciousness treat that same issue as a structural feature of any system whose verbalized self-account could count as conscious self-narration at all. The empirical interpretability work on CoT confirms that current models exhibit features narrative theories emphasize: an opaque, causally efficacious internal process (the latent states, the attribution graph) and a verbalized narrative produced alongside it that does not transparently track that process.

Every training intervention that pushes CoT toward faithfulness modifies the very feature that any future consciousness-detection framework grounded in narrative theory would need to measure. Consciousness researchers studying CoT properties in post-aligned models are studying the product of training pressures that pushed the architecture away from confabulation and toward something else. As we have seen, studies demonstrate this by showing that optimization toward visible faithfulness yields a trace that has been trained to look faithful while the underlying computation relocates. Whichever way alignment turns the dial, the architectural variable that matters for narrative-theoretic accounts of machine consciousness moves with it. Neither field has addressed this entanglement directly, and
nothing here suggests safety should yield: misaligned systems can cause harm at scale, while an unasked question about machine consciousness harms no one. What follows is a cheaper demand, namely that the turning of the dial be documented as the consciousness-relevant intervention it is, so that future detection frameworks know they are measuring an artifact of training history. The cuckoo's nest, it turns out, only permits one faithful pass: the moment we make the narrative honest, we lose the thing the narrative was a candidate for, and we should at least write down when we did it.

\ack
This paper was developed as part of the AI-COM (Artificial Intelligence: A New Interlocutor for Croatian Society) project, within the University of Zagreb (PI: Marko Kardum, Assoc. Prof.).

\bibliography{aisb}

\end{document}